\documentclass[fleqn,twoside]{article}
\usepackage{espcrc2}
\usepackage{amsmath}
\usepackage{graphicx}

\title{Lattice 2001: Reflections }
\author{
 P.~Hasenfratz\address{Institute for Theoretical Physics,
    University of Bern, Sidlerstrasse 5, CH-3012 Bern, Switzerland}} 

\begin{document}

\begin{abstract}
A few subjects which strongly intertwine our field are discussed: 
$K \rightarrow \pi \pi$ decay,
chiral symmetry on the lattice and a few other selected
topics. Open questions are touched also
on perturbation theory, locality, Gribov copies, CP
symmetry in chiral gauge theories and cut-off effects.
\end{abstract}

\maketitle

\section{Introduction}
For the final talk of this exciting conference I have
chosen a small number of subjects where the progress impressed me, or
which dominated the field lately, or simply lie close to my heart.

The first part deals with the heroic work of the CP-PACS and RBC
groups on $K \rightarrow \pi \pi$. It is exciting and somewhat
frightening to see how brains, software and hardware come
together in a highly relevant project which is then pushed to the end
(which does not necessarily mean full success).

The second part is on chiral symmetric fermions. This field went
through a rapid development during the last few years and attracted
many contributions at this conference also.

The third part contains selected topics: topological susceptibility,
some remarks on light hadron spectroscopy and a few words on cut-off
effects.

In a few cases I will raise questions, discuss issues which keep
bugging me. These parts will be separated from the main text. It might
be that some of them reflect my ignorance only. 

\section{$K \rightarrow \pi \pi$ from $K \rightarrow \pi$ and $K
\rightarrow 0$ with domain wall fermions}

There are many reasons for discussing this problem here. The $K
\rightarrow \pi \pi$ decay contains highly relevant physics, it is a
very active field of current research and a central problem of our
community since the early years of lattice calculations\cite{2.0}. 
This complex problem is related to exciting theoretical
methods and issues like operator product expansion, (quenched) chiral
perturbation theory ($(Q) \chi PT$), composite operator renormalization,
chiral symmetry, etc.

Interesting possibilities were discussed recently\cite{2.1,2.2} to
avoid the Maiani-Testa problem\cite{2.3} due to three external
particles in the $K \rightarrow \pi \pi$ amplitude. 
Preliminary results using three-point functions have
been presented in Martinelli's plenary talk\cite{2.4,2.5}. Here I will
discuss the 'reduction method' used by the CP-PACS\cite{2.6} and RBC
groups\cite{2.7}. In this case the $K \rightarrow \pi \pi$
matrixelement is related to the $K \rightarrow \pi$ and $K \rightarrow
0$ matrixelements\cite{2.8}. Without a chiral symmetric
regularization the problem is not tractable. Both groups used domain
wall fermions\cite{2.9}.

The $K \rightarrow \pi \pi$ decay exhibits two significant phenomena
of the standard model: the $\triangle I=1/2$ rule  and the mixing
induced and direct CP violation. The 
dynamics of these decays is determined by a non-trivial interplay of
strong and electroweak forces characterized by several
energy scales of very different magnitude from $m_t, m_W$ to $m_d,
m_u$. At present there is no way to treat all these scales at the same
time on the lattice. To disentangle long and short distance
contributions operator product expansion is used.

\subsection{Operator product expansion (OPE)}
\begin{figure}[ht]
\hspace{0.5cm}
\includegraphics[scale=0.40]{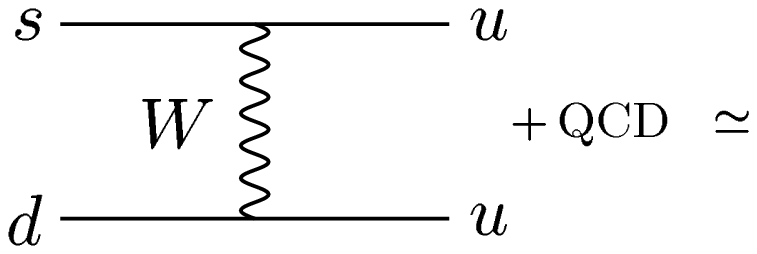}
\includegraphics[scale=0.40]{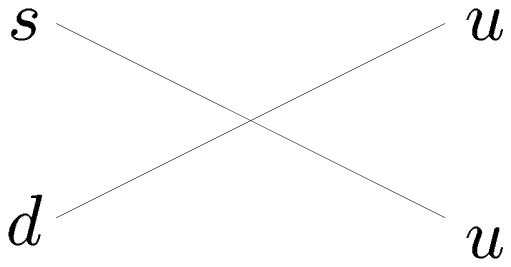}
\caption{The leading order quark level graph and the corresponding
four fermion interaction. }
\label{fig:fig1}
\end{figure}

Consider the $K^0 \rightarrow \pi^+ \pi^-$ transition in leading order
of the weak interactions. The quark level transition on the l.h.s.
of Fig.~\ref{fig:fig1} is dressed with QCD interactions (the spectator
quark is suppressed).
Since $M_W$ is large relative to the typical momentum $p$ of this problem
and so the $W$ boson propagates over a short distance, the four fermion
interaction is almost point like. This makes a systematic expansion
possible for this amplitude:
\begin{equation}
A=C(\mu/M_W,\alpha_s)\langle Q\rangle 
\end{equation}
with $O(p^2/M_W^2)$ corrections.
QCD effects above the factorization scale $\mu$ are included in the
Wilson coefficient $C$, the low energy contributions below $\mu$ are
collected into the matrixelement of the local four fermion operators
$Q$. This concept of factorization lies at the hart of many QCD
applications and I am not aware of reasons to question its applicability.

If $\mu \gg \Lambda_{QCD}$, we expect that the Wilson
coefficient $C(\mu/M_W,\alpha_s)$ can be calculated in perturbation
theory (PT). The factorization implies that $C$ is independent of the
external states and can be calculated using off-shell quarks with high
virtuality on both sides of the equation in Fig.~\ref{fig:fig1}. 
The scale $\mu$ has
to be chosen judiciously. It should be large enough to justify PT when
calculating $C$. On the other hand, if only three light quarks are
simulated on the lattice (charm is integrated out), $\mu$ can not be
larger than $m_c \approx 1.3$ GeV.

\vspace{0.5 cm}
{\small\rm
While factorization is a basic assumed feature, the question whether
the Wilson coefficients can be calculated perturbatively at a given
$\mu/\Lambda_{QCD}$ is a technical, but important issue.

One can check consistency within PT itself by considering the
convergence properties of the expansion, but this is not easy and also
not the whole story. With the lattice formulation we have a tool to
check the performance of PT by comparing it with the full
non-perturbatively calculated result. PT is used in different
situations: \\
{\bf i) Physical quantities with a high scale}\\
The number of available well measured short distance physical
quantities is very limited. The scale dependence of the running
coupling in the Schr\"odinger functional scheme
$\alpha_{SF}(\mu)$\cite{4.1} 
is in good agreement with (3-loop) PT up to
$\alpha_{SF}\approx 0.3$ in pure Yang-Mills theory\cite{4.2}. Having
the $\Lambda$-parameter from this work (in terms of the low energy
scale $r_0 \approx 0.5{\rm fm}$\cite{4.3}), PT 
in Yang-Mills theory is a parameter free expansion.
\begin{figure}[ht]
\includegraphics[scale=0.30]{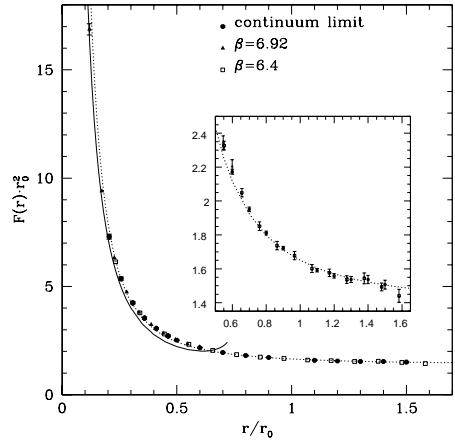}
\vspace{-0.8cm}
\caption{The measured static force compared with PT (solid line) and the
string model prediction (dotted line)\cite{4.4}.}
\label{fig:fig2}
\end{figure}
\begin{figure}[ht]
\includegraphics[scale=0.40]{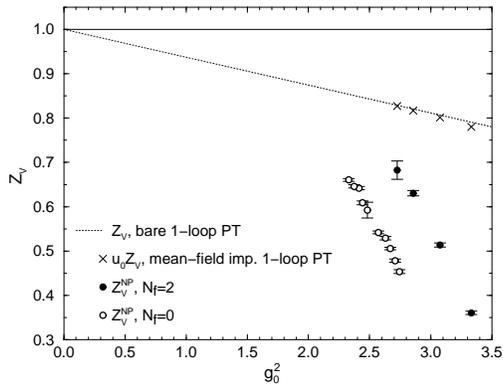}
\vspace{-0.8cm}
\caption{The $Z$-factor of the vector current\cite{6.1}.}
\label{fig:fig3}
\end{figure}
In a recent work\cite{4.4} the potential/force was measured at short
distances $0.05 {\rm fm}\le r \le 0.8 {\rm fm}$ deep in the continuum
limit and the result was compared with PT. The authors pointed out
that the perturbative scheme adopted should be chosen judiciously - a
point which sheds additional light on the troubled history of PT for
the quark potential\cite{5.1}. The scheme defined by the force
$\alpha_{qq}(\mu)=r^2 F(r)/C_F, \mu=1/r$ looks well behaving and 
describes reasonably the non-perturbative data up to $r \approx 0.15 {\rm
fm}\approx (1.3 {\rm GeV})^{-1}$. Fig.~\ref{fig:fig2}
shows another well known feature which is, however, difficult to
understand. Including the leading bosonic string model correction in
the force $F(r)= \sigma+\pi/12r^2$\cite{5.2}, which is expected to
work at large distances, describes the data well at quite short
distances also. Actually, even at $r \approx 0.15 {\rm fm}$ this form is
closer to the data than PT. The picture is confusing since, even at
the largest distance of this analysis ($1.6 {\rm fm}$) the spectrum of
the effective QCD string deviates significantly from that of an
effective bosonic string\cite{5.3}.\\
{\bf ii) Renormalization constants $Z$}\\
The scale dependent renormalization constants of composite operators
depend on the cut-off and on the renormalization point $\mu$. When
combined with renormalization group (RG) this should be a
controlled perturbative problem if these scales are large. Nevertheless,
as it is well known, PT performs poorly on the renormalization
factors. The non-perturbative \{1-loop boosted perturbative\} pseudoscalar
renormalization factor $Z_P$, for example, at $\mu=2$GeV, $1/a\approx 2$GeV
is 0.45(6)\{0.62\}\cite{5.6}, 0.39(3)\{0.59\}\cite{5.7} and
0.34\{0.54\}\cite{5.8}  
for Wilson, non-perturbatively improved Wilson and staggered actions,
respectively\cite{5.5}. 
The situation is not always better with the finite
renormalization factors, like that of the naive vector current
$Z_V$. This renormalization factor has no scale ($\mu$) dependence, it
is a function of the lattice coupling $g_0$ and goes to 1 in the
continuum limit. 
Fig.~\ref{fig:fig3}, taken from\cite{6.1} \footnote{I
am indebted to Tomoteru Yoshie for a correspondence concerning the
results on $Z_V$ 
in\cite{6.1} and\cite{6.11}.}, contains quenched $a \in 0.11-0.20 {\rm fm}$
and $N_f=2$ $a \in 0.11-0.23 {\rm fm}$ data (RG improved gauge action
\cite{6.9}, mean field improved clover quark action) compared with bare and
mean-field improved PT.\\
{\bf iii) Cut-off effects, improvement coefficients}\\
In order to eliminate the leading cut-off effects (in spectral
quantities) the action is extended by a term $C_{SW}(g_0) Q$, where
$Q$ is the Sheikholeslami-Wohlert operator\cite{6.2}, while $C_{SW}$
is the Symanzik improvement coefficient\cite{6.3}. The improvement
should work for any external states, also for those with high
virtuality, which can be used to calculate the improvement
coefficients $C_{SW}(g_0)$. (Notice the analogies with OPE.) For
cut-off values of present day simulations ($1/a \le 3$GeV) PT does not
reproduce the non-perturbatively fixed $C_{SW}(g_0)$\cite{6.4}. For
further examples of problems with PT when calculating operator
improvement coefficients see Table I in\cite{6.5}.

The conclusion is that in the very few cases where PT was
systematically compared with numerical results on physical
quantities a judiciously chosen PT scheme 
seems to work well down to such low
scales as $\mu \approx 1.3$GeV. On the other hand, in problems where
the cut-off also plays a role PT performs poorly for $1/a \le 3$GeV.
Many believe that this is related to some special feature of
lattice PT, like the existence of tadpole graphs. I do not think that
this is the full explanation. First, a lot of effort was invested  to
overcome this problem of lattice PT\cite{6.6}. Second, more
importantly, some of the quantities in question, do not look like
perturbative at all. The $a$ dependence of the coupling $g_0$ of the
plaquette action (lattice $\beta$-function)\cite{6.7}, the $g_0$
dependence of $C_{SW}$\cite{6.4}, or $Z_V$ in Fig.~\ref{fig:fig3},
for example, simply do not invite to use PT at all.
}

\vspace{0.5 cm}
The graph in Fig.~\ref{fig:fig1} actually generates two four-fermion
operators with different color structure
\begin{align}
&Q_1=({\bar s_a}u_b)_L({\bar u_b} d_a)_L \,,\\
&Q_2=({\bar s_a}u_a)_L({\bar u_b} d_b)_L \,,\nonumber 
\end{align}
where $a,b$ are color indices and the short-hand notation 
$({\bar s_a}u_b)_{R/L}={\bar s_a}\gamma_\mu(1\pm \gamma_5)u_b$ is used
here and below. The sum of these operators (weighted with the Wilson
coefficients which include $V_{us}V_{ud}^*$ from the CKM matrix)
taken between the incoming kaon and the outgoing pions in isospin
$I=0$, or 2 states
\begin{equation}
\hspace{-0.5cm}
K^0 \rightarrow (\pi \pi)_I\,=\,A_I\exp(i\delta_I)\,=
\label{7.2}
\end{equation}
\vspace{-1.0cm}
\begin{equation}
\hspace{-0.5cm}
\frac{G_F}{\sqrt 2}V_{us}V_{ud}^*\sum_{i=1}^2 z_i(\mu)
\langle \pi \pi_I|Q_i|K\rangle^{\overline {MS}}(\mu), \nonumber
\end{equation}
should explain the large ratio
\begin{equation}
\frac{1}{\omega} = \frac{{\rm Re}(A_0)}{{\rm Re}(A_2)}\approx 22 \,.
\end{equation}
i.e. the $\triangle = 1/2$ rule. In eq~(\ref{7.2})
$\delta_I$ is the phase shift from the final state $\pi \pi$
interaction, and $z_i(\mu) \propto C_i(\mu)$ Wilson coefficients (and
the additional ones for $i=3,\dots,10$ below) are known in the
standard model at next-to-leading order\cite{7.1}. On the other hand,
the weak interaction part of the 'tree graph' in Fig.~\ref{fig:fig1}
does not know about the third generation and is so CP conserving. In
order to see CP violation we have to go beyond this leading
approximation and include the QCD penguin and electroweak diagrams of 
Fig.~\ref{fig:fig4}. 
\vspace{-0.8cm}
\begin{figure}[ht]
\hspace{3.5cm}
\includegraphics[scale=0.40]{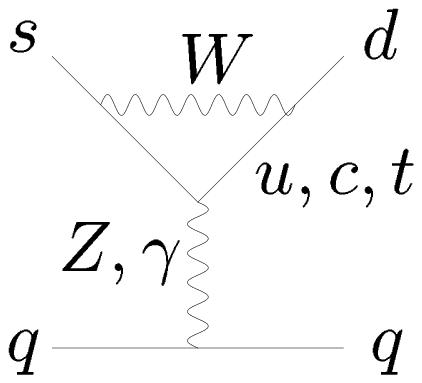}
\end{figure}
\vspace{-5.1cm}
\begin{figure}[ht]
\hspace{1.0cm}
\includegraphics[scale=0.40]{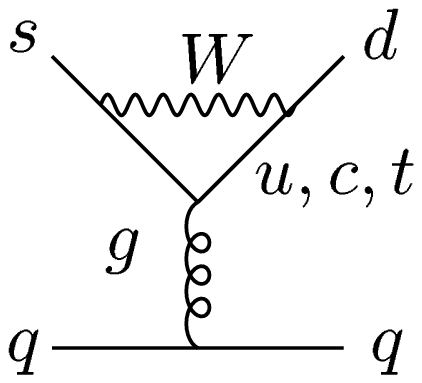}
\vspace{-2.1cm}
\caption{The QCD and electroweak penguins.}
\label{fig:fig4}
\end{figure}
The resulting effective hamiltonian is a
sum over 10 operators\cite{7.1}. The $\triangle = 1/2$ problem is
expected to be dominated by the $Q_1$ and $Q_2$ contributions since
the functions $z_i(\mu=m_c)$ are small for $i=3,\dots,10$
\cite{7.1}. On the other hand, phenomenological considerations
suggest\cite{2.0} that the imaginary part of the amplitude 
\begin{equation}
\hspace{-0.2cm}
{\rm Im} A_I=
\end{equation}
\vspace{-1.0cm}
\begin{equation}
\hspace{-0.2cm}
-{\rm Im}(V_{ts}V_{td}^*)\frac{G_F}{\sqrt(2)}\sum_{i=3}^{10}
y_i(\mu)\langle Q_i\rangle^{\overline {MS}}_I(\mu)\,, \nonumber
\end{equation}
where $\langle Q_i\rangle_I\exp(i\delta_I)=\langle \pi \pi_I|Q_i|K\rangle $, is dominated by
$\langle Q_6\rangle_{I=0}$ and $\langle Q_8\rangle_{I=2}$. In this approximation the direct CP violation
is given by
\vspace{-0.3cm}
\begin{equation}
\hspace{-0.3cm}
\frac{\epsilon'}{\epsilon}\approx \frac{\omega G_F}{2|\epsilon|{\rm Re}A_0}
{\rm Im}(V_{ts}V_{td}^*)\times
\label{8.3} 
\end{equation}
\vspace{-0.5cm}
\begin{equation}
\hspace{-0.3cm}
\left(y_6(\mu)\langle Q_6\rangle^{\overline {MS}}_{I=0}(\mu)-\frac{1}{\omega}
y_8(\mu)\langle Q_8\rangle^{\overline {MS}}_{I=2}(\mu)\right) \nonumber
\end{equation}
where
\begin{align}
\label{8.4}
&Q_6=({\bar s}d)_L \left(
({\bar u} u)_R +({\bar d} d)_R+({\bar s} s)_R\right)\,,\\
&Q_8=\frac{1}{2}({\bar s}d)_L \left(
(2{\bar u} u)_R -({\bar d} d)_R-({\bar s} s)_R\right)\,.\nonumber
\end{align}
It is reassuring that this qualitative expectation is 
supported by the numerical data\cite{2.6,2.7}.
Although the contribution from $Q_8$ is
suppressed by the electroweak coupling constant, this is largely
compensated by the prefactor $1/\omega$ leading to a numerically unpleasant
cancellation between the two terms in eq.~(\ref{8.3}).

\subsection{Determination of the bare \\
$\langle \pi \pi_I|Q_i|K^0\rangle $ matrixelements}
The CP-PACS and RBC collaborations followed the 'reduction method',
where tree level $\chi PT$ is used to relate the $K^0 \rightarrow \pi
\pi$ matrixelements to those of $K^+ \rightarrow \pi^+$ and $K
\rightarrow$ vacuum\cite{2.8} which are then calculated on the
lattice. 

Although the reduction step simplifies the problem significantly, it
introduces an unpleasant mixing with a lower dimensional two-quark
operator. The $(8_L,1_R)$ operators, like $Q_6$ in eq.~(\ref{8.4}),
mix with the two-quark operator $'{\bar s}d\,'$ since the singlet right
handed part communicates with the vacuum\cite{2.8}.
This $(8_L,1_R)$ operator should be CPS
(CP + $d \leftrightarrow s$) invariant (as all the $Q_i$ operators
are) leading to the form for $m_s \ne m_d$
\begin{equation}
\hspace{-0.5cm}
Q_{sub}=(m_s+m_d){\bar s}d - (m_s-m_d){\bar s}\gamma_5 d 
\label{9.1} 
\end{equation}
\vspace{-0.9cm}
\begin{equation}
\hspace{-0.5cm}
=\partial_\mu \left(\frac{m_s+m_d}{m_s-m_d}
{\bar s}\gamma_\mu d - \frac{m_s-m_d}{m_s+m_d}
{\bar s}\gamma_\mu \gamma_5 d \right)\,.\nonumber
\end{equation}
In eq.~(\ref{9.1}) the equation of motion was used to write $Q_{sub}$
as a total divergence for $m_s \ne m_d$. In the physical 
$K^0 \rightarrow \pi \pi$ matrixelement this total divergence does not
contribute since the weak operator carries zero momentum. On the other
hand, for the $K^+\rightarrow \pi^+$ reduced matrixelement $Q_{sub}$
gives a (quadratically divergent) contribution: if $m_s=m_d$
$(m_\pi=m_K)$ is used in the simulation (CP-PACS,\cite{2.6}) then
$Q_{sub}$ is not a total divergence; if $m_s \ne m_d$ (RBC,
\cite{2.7}) then the operator carries momentum and the total
divergence contributes. This quadratic divergence should be subtracted
from the $\triangle I =1/2$ operator matrixelements with a coefficient
determined by the corresponding $K \rightarrow 0$ matrixelement. The
reduction equations have the form ($f_\pi \approx 93$ MeV)
\begin{equation}
\hspace{-0.5cm}
\langle \pi^+ \pi^-|Q^{\triangle I=1/2}_i|K^0\rangle =
\label{10.1}
\end{equation}
\vspace{-0.5cm}
\begin{equation}
\hspace{-0.5cm}
i\frac{m_K^2-m_\pi^2}{\sqrt 2 f_\pi M^2} 
\langle \pi^+|Q^{\triangle I=1/2}_i-\alpha_iQ_{sub}|K^+\rangle \,,\nonumber
\end{equation}
for $i=1,\dots,6,9,10$ and $\alpha_i$ is determined by
\begin{equation}
\langle 0|Q^{\triangle I=1/2}_i-\alpha_i Q_{sub}|K^+\rangle =0 \,.
\end{equation}
The $\triangle I=3/2$ operators have no subtractions and
for $i=1,\dots,6,9,10$ the relations are like in  eq.~(\ref{10.1})
with $\alpha_i=0$. For $i=7,8$ we have
\begin{equation}
\langle \pi^+ \pi^-|Q^I_i|K^0\rangle =-\frac{1}{\sqrt 2 f_\pi}
\langle \pi^+|Q^I_i|K^+\rangle \,. \nonumber
\end{equation}
In the equations above $m_K$ and $m_\pi$ are the physical masses,
while in the $K \rightarrow \pi$ matrixelement $m_K=m_\pi=M$. Remarks:

\noindent
$\bullet$ The reduction is on the tree level of (Q)$\chi$PT only. The
$O(p^4)$ low 
energy constants needed for $K \rightarrow \pi \pi$ can not be found
from the $K \rightarrow \pi$, $K \rightarrow 0$ matrixelements. One
can use the $O(p^4)$ (Q)$\chi$PT results for $K \rightarrow \pi$
\cite{10.1} to make a more reliable fit for the tree level constants,
but even in this case $K \rightarrow \pi \pi$ will be obtained on the
tree level only. The 1-loop chiral corrections are estimated to be
significant ($m_K \approx 500$MeV !)\cite{10.1,10.2}.

\noindent
$\bullet$ The reduction step relies heavily on chiral symmetry. The mixing
problem can not be treated in the $\triangle I=1/2$ sector with Wilson
fermions. 

\noindent
$\bullet$ Both groups demonstrated that a controlled signal can be obtained
after the power divergent subtraction.

\noindent
$\bullet$ As eq.~(\ref{10.1}) shows the $K^+ \rightarrow \pi^+$ matrixelements
should vanish in the chiral limit $M \rightarrow 0$ for $i \ne
7,8$. CP-PACS finds that this is the case within the errors, RBC
observes small non-zero intercepts which might be related to the
residual chiral symmetry breaking of the domain wall
fermions. As eq.~(\ref{10.1}) shows, the slope is all what one needs. 

\subsection{Renormalization}
CP-PACS used renormalization factors calculated in 1-loop PT
\cite{11.1}. RBC applied the Roma-Southampton
non-perturbative method with gauge fixing\cite{11.2}.

\vspace{0.5 cm}
{\small\rm
The uncontrolled effect of Gribov copies in the gauge fixing is a
frequently discussed issue in this non-perturbative renormalization
method. Undoubtedly, it would be very difficult to perform an unbiased
averaging over the Gribov copies\cite{11.3}. On the other hand, is
such averaging really needed if we study a short distance ($\propto
1/\mu$) problem? Is it consistent to worry about Gribov copies and, at
the same time accepting the validity of PT at the scale $\mu$ (say, in
$\overline {MS}$), which knows nothing about such non-perturbative
features? 
}
\subsection{Physical results}
The parameters of the simulations of CP-PACS and RBC are quite similar.
On the real part of $A_I$ and so on the $\triangle I=1/2$ rule, RBC
obtained results close to those in experiments, while the results on
${\rm Re}A_0$ and $\omega^{-1}$ from CP-PACS are a factor of $\sim 2$
smaller.

Both groups obtained a small negative $O(10^{-4})$ number for
$\epsilon'/\epsilon$ as opposed to the experimental average of $(17.2
\pm 1.8)10^{-4}$\cite{12.1}.

Given the large theoretical and numerical complexity of the problem I
find these works very impressive. There are several possible sources
of systematical error, like quenching, large corrections to tree level
reduction in $\chi$PT and (continuum) PT at scales $\sim 1.3$GeV. 
In\cite{12.2} the authors argue that a new low energy constant (a
pure quenching artifact) enters the reduction relations which might
have an influence on the results above.

\section{Chiral symmetric lattice fermions}
We had four plenary talks at this conference which were directly
on this subject\cite{13.1}, or were closely related\cite{13.2}. I
would like to discuss here a few points only\cite{13.0}.

The Ginsparg-Wilson (GW) relation\cite{13.3} was suggested as the
'mildest way' to break chiral symmetry on the lattice in 1982
immediately after the no-go theorem of Nielsen and Ninomiya
\cite{13.4}. In its simplest form the GW relation reads
\begin{equation}
\hspace{-0.3cm}
\gamma_5 D + D \gamma_5 = D \gamma_5 D \,,
\label{13.1}
\end{equation}
or, equivalently (if $D$ has no zero modes)
\begin{equation}
\hspace{-0.3cm}
\gamma_5 D^{-1}(x,x') + D^{-1}(x,x') \gamma_5 = \gamma_5 \delta_{x,x'} \,.
\end{equation}
Although the propagator $D^{-1}$ does not anticommute with $\gamma_5$
the violation is a contact $\sim \delta_{x,x'}$ term only. It is
expected and is really so that the GW relation implies chiral symmetry
on physical predictions\cite{14.1}.

Eq.~(\ref{13.1}) is a non-linear relation for $D$. It is clear
intuitively and has been shown rigorously\cite{14.2} that the
solution can not be ultralocal, it has a tail. On the other
hand, a physically interesting solution should be {\it local}, which
means that this tail of $D$ should decay faster than the signal at
physical distances in correlation functions
\begin{equation}
\hspace{-0.4cm}
D(x,x') \sim \exp{(-M a |x-x'|)}_{|x-x'|\gg 1} \,,
\end{equation}
where $M=O$(cut-off).

\vspace{0.5 cm}
{\small\rm
Let me discuss the important issue of locality of the action in a QFT
further. Locality is related to universality and universality implies
predictive power. If,
however, the action has couplings between distant points which compete
with the real dynamics of the system, then the predictions will depend
on the non-physical microscopical details. Universality will be lost
and such an action is not useful.

For this reason, the checks concerning the locality of the GW type
Dirac operators are very important
\cite{13.1,15.1,15.2,15.3,15.4}. Dynamical staggered fermions with $N_f <
4$ present a not fully understood, potentially dangerous situation. It
is not obvious that $({\rm det}D(U)_{st})^{1/2}$ in the path
integral defines a local theory, although there are no arguments
excluding this either\cite{15.5}. Two dimensional models might be the right
place for testing.
}

\vspace{0.5 cm}
Since no solution to the GW relation eq.~(\ref{13.1}) was found in the
presence of gauge fields, the idea was soon after its birth
abandoned. The first lattice regularization of fermions with chiral
symmetry, the domain wall fermions\cite{2.9} and the related overlap
construction\cite{15.6} followed a different path and seemed to be
unrelated to the GW relation.

Following the observation that the fixed-point Dirac operator
satisfies the GW relation\cite{15.7} the interest turned to this
general formulation again. The GW relation is a powerful theoretical
tool. Its immediate consequence is the index theorem on the lattice
\cite{15.8} and it implies the existence of an exact chiral symmetry
transformation\cite{15.9}. The observation that the overlap Dirac
operator satisfies the GW relation connected the domain wall approach
to this general formulation\cite{15.10}.

As the previously discussed $K \rightarrow \pi \pi$ studies
illustrate, domain wall and, to a lesser extent, overlap fermions are
used today in large scale quenched simulations in problems, where
chiral symmetry plays an important role\cite{16.0,2.6,2.7}.

The fixed-point gauge and Dirac actions are defined by classical
equations\cite{16.2}. These equations can be solved numerically in an
iterative way which is, however, far too expensive in a stochastic
calculation. A parametrization in terms of a finite number of
operators is unavoidable. This can be done quite effectively and the Dirac
operator obtained this way performs well in different test runs
\cite{16.3,15.3}. An alternative way to fix the couplings in a
parametrized Dirac operator is a systematic expansion of the GW
equation itself\cite{16.5}. For test results on this action
and for comparison with other formulations, see\cite{16.7}. 

The new chiral symmetric actions allow to investigate the role of
topological excitations in the QCD vacuum. It was convincingly
demonstrated in several contributions at this conference that an
instanton dominated picture of low eigenmodes of the Dirac operator is
consistent: the local peaks in these eigenmodes are dominantly chiral
\cite{16.6,16.7}. Such a study is hardly possible with Wilson fermions
\cite{16.8}. 

Chiral symmetric Dirac operators have nice theoretical properties
including the absence of exceptional configurations and $O(a)$
improvement in spectroscopy. They are, however, expensive to
simulate. The first studies on spectroscopy are exploratory
\cite{16.0,16.10,16.11,16.12,16.3}. From these works it is obvious that light
hadron spectroscopy becomes much cleaner: quark mass tuning and long
chiral extrapolations are not needed, topology is well defined and
operator renormalization is significantly simplified. The first
quantitative results confirm that there are no
unexpected, hidden problems with this regularization of
QCD. Simulations in the Schwinger model\cite{16.13} strengthen
this conclusion.

There are problems, like the fermion condensate, where a chiral
symmetric action is a prerequisite for a quantitative study. The quark
condensate, or more precisely, the leading low energy constant
$\Sigma$ in the chiral Lagrangian, has been calculated by several
groups using different actions and methods\cite{17.1,16.11,15.3}. 
The results are consistent and scatter
around $(\Sigma^{\overline {MS}}(2{\rm GeV}))^{1/3}=270 {\rm MeV}$ with a
few percent statistical error. Since no continuum extrapolation is
done yet, the systematical error is unknown.

As mentioned before, eq.~(\ref{13.1}) implies exact symmetry under the
transformation\cite{15.9}
\begin{align}
\label{17.1}
&\psi \rightarrow \psi + i \epsilon T^a \gamma_5 (1-D) \psi \,,\\
&{\overline {\psi}} \rightarrow {\overline {\psi}} + i \epsilon 
{\overline {\psi}} \gamma_5 T^a \,,\nonumber
\end{align}
$a=1,\dots,N_f^2-1$, which is a sort of smeared $\gamma_5$ transformation.
The action is also invariant under a singlet chiral transformation,
but the fermion measure is not (notice that the transformation in
eq.~(\ref{17.1}) is gauge field dependent) leading to the correct
anomalous Ward identity
\begin{equation}
\langle \delta {\cal O}\rangle_F -2 N_f\nu(U) \langle {\cal O}\rangle_F = 0\,,
\end{equation}
where $\nu(U)$ is the topological charge, while $\langle ~.~\rangle_F$ is the fermionic
path integral on a fixed gauge field configuration $U$.

The transformation in eq.~(\ref{17.1}) is asymmetric on $\psi$ and
${\overline {\psi}}$, and so are the left/right projectors also
\cite{13.0}. 
\begin{equation}
\hspace{-0.4cm}
\psi_{L/R}={\hat P}_{L/R} \psi\,,\qquad 
{\overline {\psi}}_{L/R} = {\overline {\psi}}P_{R/L} \,,
\label{17.3}
\end{equation}
where
\begin{equation}
\hspace{-0.4cm}
{\hat P}_{R/L}=\frac{1}{2}(1\pm {\hat \gamma_5})\,,\qquad 
P_{R/L}=\frac{1}{2}(1\pm \gamma_5)\,, \nonumber
\end{equation}
\vspace{-0.5cm}
\begin{equation}
\hspace{-0.4cm}
{\hat \gamma_5}=\gamma_5(1-D).
\label{18.1}
\end{equation}
The fermion action falls into $L$ and $R$ parts as in the formal continuum 
\begin{equation}
\hspace{-0.4cm}
{\overline {\psi}}D\psi={\overline {\psi}}_L D\psi_L+
{\overline {\psi}}_R D\psi_R \,.
\label{18.2}
\end{equation}
Eqs.~(\ref{17.3},\ref{18.1},\ref{18.2}) are the first, almost trivial
steps in a highly 
complex construction which led to a breakthrough concerning chiral
gauge theories\cite{18.0}. For anomaly free complex representations
for $U(1)$, or $SU(2)\times U(1)$ gauge groups the theory is
constructed non-perturbatively\cite{18.1,18.2}, for a general compact
group it is constructed to all orders of PT\cite{18.3}.

At present, chiral gauge theories obtained this way have a certain ambiguity
which I would like to discuss briefly. A normalization factor (absolute value and
phase) in $Z_\nu$, where $Z_\nu$ is the partition function of the
theory in the topological sector with charge $\nu$, is left undetermined. This
is an unsatisfactory situation since, among others, it leaves the magnitude of
fermion violating processes undetermined.

In a recent paper\cite{18.4} H. Suzuki considered Weyl fermions in a real
representation of the gauge group. He demonstrated that this Weyl fermion is
equivalent to a Mayorana ($L-R$ symmetric) fermion on the lattice, like in the
continuum. Based on this observation Suzuki suggested a 'natural' relative
normalization of different topological sectors for Weyl fermions in 
{\it complex} representations. 

\vspace{0.5 cm}
{\small\rm
The normalization of different topological sectors is related to CP symmetry
in chiral gauge theories - a problem which keeps bugging me since 1998
\footnote{Martin L\"uscher raised my attention
to the fact that the projection operators ${\hat P}_{R/L}$ in eq.~(\ref{18.1})
do not follow the continuum transformation laws under CP.}. The present
formulation is not explicitly CP symmetric. It might turn out that the
renormalized final physical predictions, which include the relative
normalizations between different topological sectors,
respect CP. If this is the case, the
existence of an explicit CP invariant lattice formulation is not 
excluded and would be interesting to find. If CP symmetry can not be restored
(CP anomaly), that would be even more interesting. 

Under CP, $\psi \rightarrow {\overline {\psi}}W\,, {\overline {\psi}}
\rightarrow -W^{-1}\psi\,, U \rightarrow U^{\rm CP}$, the vector gauge theory is
invariant if $W D(U^{\rm CP})^T W^{-1} = D(U)$, where $U^{\rm CP}$ is the
CP-transformed gauge configuration and in my convention
$W=\gamma_2$.\footnote{Under CP the spatial coordinates are
reflected which we keep implicit in the following equations.} I
assume that $D$ satisfies this condition. However, unlike in the
continuum, the left handed part of the fermion action is not invariant:
\begin{equation}
{\overline {\psi}}_L D \psi_L \stackrel{{\rm CP}}{\longrightarrow} \,\,\,
{\overline {\psi}}_L D \psi_L-\frac{1}{2}{\overline
{\psi}}D\gamma_5D\psi\,.
\label{19.1}
\end{equation}
One might try to solve this problem by observing that the choice for
the projection operators in eq.~(\ref{18.1}) is not unique. One can
find a
continuous set of projectors 
constructed in terms of $\Gamma_5^{(s)}=\gamma_5(1-sD)/{\cal N}^{(s)}$
and ${\overline {\Gamma_5^{(s)}}}= (1-(1-s)D)\gamma_5/{\cal N}^{(s)}$
for the fermions and antifermions, respectively, where $s$ is an
arbitrary real parameter and ${\cal N}^{(s)}$ is for the correct
normalization. These projectors
assure the $L+R$ decomposition of eq.~(\ref{18.2}) and give $\nu$ for
the difference between the dimension of the left handed fermion and
antifermion 
spaces\footnote{This is the source of fermion number violation.} for
any value of $s$.
On the other hand, the CP violating term in  eq.~(\ref{19.1}) will be
proportional to $1-2s$ and so for $s=1/2$ the
left handed action becomes CP invariant. Unfortunately, just at this
value of $s$ the projectors become singular (non-local). It seems to
be difficult to define a CP invariant left handed action.

Return back to the original definition in 
eq.~(\ref{18.1}) and consider the fermionic expectation value on the
gauge field configuration $U$
\begin{equation}
\hspace{-0.9cm}
\langle {\cal O}\rangle_F=K_\nu \int D\psi_L D{\overline \psi}_L 
\exp(-{\overline \psi}_L D(U) \psi_L)\, {\cal O}\,,
\label{20.1}   
\end{equation} 
where ${\cal O}$ is some expression of the left handed fields and
$K_\nu$ is the unknown normalization factor of the topological
sector $\nu$. Introducing a basis for the fermions and antifermions
\begin{equation}
{\hat P}_Lv_j=v_j\,,\qquad \psi_L(x)=\sum_j c_jv_j(x)\,,
\label{20.2}
\end{equation}
\vspace{-0.3cm}
\begin{equation}
P_R \omega_k=\omega_k\,, \qquad {\overline \psi}_L(x)=
\sum_k{\overline c}_k \omega^\dagger_k \,, \nonumber
\end{equation}
eq.~(\ref{20.1}) can be written as
\begin{equation}
\langle {\cal O}\rangle_F=K_\nu \prod_{j,k} \int dc_j\int d{\overline c}_k 
\label{20.3}   
\end{equation}
\vspace{-0.3cm}
\begin{equation}
\hspace{1.0cm} 
\exp(-\sum_{k,j}{\overline c}_k M_{k,j} c_j)\, {\cal O}\,,\nonumber
\end{equation}
where $M_{k,j}=(\omega_k,Dv_j)$. Going over to a different basis, the
measure changes by a gauge field dependent phase. The difficult part
of defining a chiral gauge theory is to fix this phase (or the basis)
so that the final theory is local and gauge invariant
\cite{18.0,18.1,18.2,18.3}. 

The absolute value $|\langle {\cal O}\rangle_F|$ is, however, independent of the
basis chosen. One might construct the $\{v_j\}$ and $\{\omega_k\}$
basis vectors in eq.~(\ref{20.2}) in terms of the eigenvectors of $D$,
for example.

Consider a gauge field configuration $U$ ($U^{\rm CP}$)
with topological charge $\nu=1$ ($\nu=-1$). 
$D(U)$ has two left
handed fermions with $\lambda=0$ and 2, further a $\lambda=2$ left handed
antifermion. $D(U^{\rm CP})$ has a left handed
antifermion with $\lambda=0$ and no left handed modes with
$\lambda=2$. The complex eigenvalues on $U$ and $U^{\rm CP}$ are the
same. 

Let us chose a simple operator ${\cal O}=c_0=(u_0,\psi)$, where $u_0$
is the zero mode $D(U)u_0=0$. Eq.~(\ref{20.3}) gives on the gauge
field $U$
\begin{equation}
|\langle {\cal O}\rangle_F|=|\langle c_0\rangle_F|=
|K_{+1}|\,2\,\prod |\lambda|\,,
\label{21.1}
\end{equation}
where every complex pair of eigenvalues $\lambda, \lambda^*$ enters
the product once. The factor 2 comes from the $\lambda=2$ left handed
mode. 

Compare eq.~(\ref{21.1}) with $|\langle {\overline {\cal {O}}}\rangle_F|$ on the
configuration $U^{\rm CP}$, where ${\overline {\cal O}}={\cal O}^{\rm
CP}=({\overline \psi},s_0)$ and $s_0$ is the right handed zero mode on 
$U^{\rm CP}$. One obtains  
\begin{equation}
\hspace{-0.1cm} 
|\langle {\cal O}^{\rm CP}\rangle_F|=
|K_{-1}|\,\prod |\lambda|\,.
\label{21.2}
\end{equation}
In eq.~(\ref{21.2}) the factor of 2 is missing: on $U^{\rm CP}$ 
with $\nu=-1$ there is no left handed mode with $\lambda=2$.

Suzuki, using a consideration unrelated to CP symmetry, suggested a
'natural' relative normalization $K_\nu=2^{-\nu/2}$
\cite{18.4}. This factor would restore CP symmetry in the absolute
value of the expectation values above. The question remains, however,
whether this choice is consistent with cluster decomposition in the
case where an instanton and anti-instanton are separated in
a $\nu=0$ configuration.
}

\section{Selected topics}

\subsection{The topological susceptibility}
The topological susceptibility
\begin{equation}
\chi = \langle  \int d^4x q(x) q(0) \rangle = \frac{\langle \nu^2\rangle}{V}\,,
\label{23.0}
\end{equation}
where $q(x)$ is the topological charge density and $\nu=\int d^4x
q(x)$. This quantity is related to one of the fanciest features of
QCD, like the $U(1)$ problem, the axial anomaly and their interplay
with topology.

The topological susceptibility $\chi^q$ calculated in quenched QCD
(i.e. over configurations generated with the Yang-Mills action) enters
the Witten-Veneziano relation\cite{23.1}
\begin{equation}
\chi^q = \frac{f_\pi^2 m_{\eta'}^2}{2 N_f} \,,\qquad N_c \rightarrow \infty
\label{23.1}
\end{equation}
in the chiral limit. In full QCD, configurations with non-zero
topological charge $\nu$ are suppressed in the chiral limit by the
fermion determinant. The leading correction to the chiral limit reads
\cite{23.2,23.3} 
\begin{equation}
\chi = \frac{m_q \Sigma}{N_f}=\frac{f_\pi^2 m_\pi^2}{2 N_f} \,,
\label{23.2}
\end{equation}
where $-\Sigma = \langle {\overline u}u\rangle = \langle {\overline d}d\rangle
\dots$ is the 
quark condensate (in our normalization $f_\pi \approx 93$MeV in Nature).
For a recent brief review on the history around these relations, see
\cite{23.4}, while for a recent summary on numerical simulations of
$\chi^q$ I refer to\cite{23.5}. Here I would like to mention some
recent developments only.

On the theoretical side the GW formulation (which has exact
$SU(N_f)\times SU(N_f)$ symmetry with the proper $U(1)$ anomaly
and the index theorem) allows a clean derivation of 
eqs.~(\ref{23.1},\ref{23.2}) on the lattice. Under the standard dynamical
assumptions (chiral symmetry is spontaneously broken and the $\eta'$
remains heavy) eq.~(\ref{23.2}) was derived in\cite{23.3}. In a
recent paper the Witten-Veneziano formula was also reproduced on the
lattice\cite{23.4}. It is interesting to notice the analogies between
the form of eqs.~(\ref{23.1},\ref{23.2}) and also in the steps of
derivation. Although the $\eta'$ mass is due to a hard symmetry
breaking (anomaly), it is switched off by the $N_c \rightarrow \infty$
limit similarly as the soft breaking $m_q$ is switched off in the
chiral limit.
\begin{figure}[ht]
\includegraphics[scale=0.35]{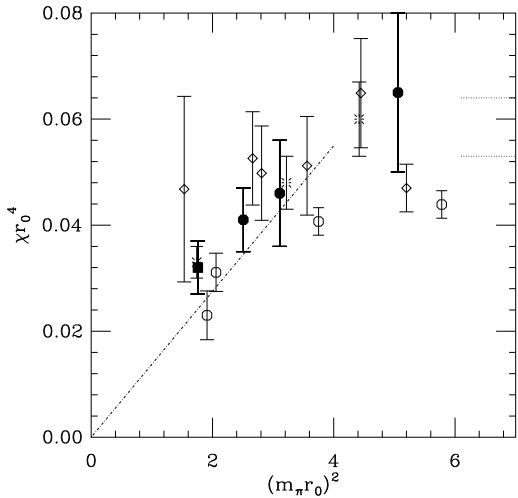}
\vspace{-0.8cm}
\caption{$N_f=2$ data: Wilson fermions\cite{25.2}(open diamond),
clover fermions\cite{25.1}(open circle), staggered fermions with
smeared link operator\cite{25.5}(filled symbols), clover fermions with
RG improved action\cite{25.7}(burst).}
\label{fig:fig5}
\end{figure}

On the numerical side, the quenched topological susceptibility seems
to settle down at a value $\chi^q=(0.205\,{\rm MeV})^4$ with an
estimated $10\%$ error on $\chi^q$\cite{24.1,23.5}. A specialty of the 
topological susceptibility is that its cut-off effects depend not
only on the action but also on the details of the calculation (type
and number of cooling steps) if non-chiral symmetric actions with a
'field theoretic' definition of $\nu$ are used. The value above
compares reasonably with eq.~(\ref{23.1}) if $N_c=3$ numbers are
inserted on the r.h.s.. $N_c=3$ seems to be close to the $N_c
\rightarrow \infty$ limit which is also supported by a recent
numerical study of the string tension, glue ball masses and $\chi^q$
in $SU(N), 2\le N \le 5$\cite{24.2}. The numerical analysis becomes
cleaner if chiral symmetric actions are simulated. Only preliminary
results are available\cite{24.3,15.3}.

The topological susceptibility of QCD is, presumably the best place to
see, how sea quarks turn Yang-Mills gauge theory in full QCD. As 
eq.~(\ref{23.2}) shows, the effect should be striking: the
susceptibility $\chi$ goes to zero in the chiral limit. The early
results\cite{25.1,25.2,25.3} appear controversial. The authors in
ref.\cite{25.5,25.51} argue that the cut-off effects are responsible
for the confusing results, in particular, chiral symmetry violation
for Wilson type fermions and flavor symmetry violation for staggered
fermions\cite{25.5} at large lattice spacings. The situation is
expected to become better at smaller $a$, with improved actions, or 
improved topological charge operators.
This expectation is
confirmed by the $N_f=2$ CP-PACS\cite{25.7} and by the
results in Fig.~\ref{fig:fig5}.

Fig.~\ref{fig:fig5} is a compilation\cite{25.5} 
of staggered, Wilson and clover
data with $N_f=2$ at $a\approx 0.1 {\rm fm}$, where, with the exception 
of the SESAM data\cite{25.2}, either the action, or the operator is
improved.  The
effect of dynamical fermions is seen clearly and the data are
consistent with $\chi \rightarrow 0$ as $m_\pi \rightarrow 0$. On the
other hand, remaining discretization errors might be significant and
the  behavior of $\chi$ with respect to $(m_\pi r_0)^2$ might look
differently in the continuum limit\cite{25.6}.

\subsection{On quenched spectroscopy}
Quenched QCD is not a healthy QFT, but believed to be universal in the
continuum limit. The data are not very convincing, but at least do not
contradict this expectation.

\vspace{0.5 cm}
{\small\rm
In my talk I referred to Sinya Aoki's contribution in Bangalore
\cite{26.1}, where the strong disagreement between the staggered and
Wilson Edinburgh plots was discussed. The situation
concerning the hyperfine splittings confused me also. By now I
understand these points better.\\
\noindent
{\it 'The staggered and Wilson fermion Edinburgh plots strongly
disagree'.}
I believe, the reason of the disagreement in Fig.4 in\cite{26.1} are
the very large cut-off effects in the unimproved staggered data which
made a continuum extrapolation unreliable. This is nicely demonstrated
in Fig.19 of ref.\cite{26.2} which shows that by reducing the cut-off
effects in the staggered simulations (by simulating at small $a$, like
in\cite{26.3}, or by using improved fermions) the
staggered Edinburgh plot is pushed towards to that obtained with
Wilson type of fermions.\\
\noindent
{\it 'Are the quenched hyperfine splittings too small, or too large?'}
They were found too small by many groups and most convincingly by the
CP-PACS Collaboration\cite{26.50}. On the other hand, UKQCD finds the
hyperfine splittings always far too large\cite{26.5}. I think the
reason is that UKQCD fixed the scale from $r_0 \approx 0.5{\rm fm}$,
while the standard choice is fixing $m_\rho$. Fixing $m_\rho$ gives
$r_0 \approx 0.55{\rm fm}$ (see, for example\cite{6.1}) and this
$10\%$ effect explains the paradox with the hyperfine splittings. The
disagreement of quenched spectroscopy with Nature can be shifted to
different quantities by changing the method of fixing the scale.
}

\subsection{Strange cut-off effects}
\begin{figure}[ht]
\includegraphics[scale=0.40]{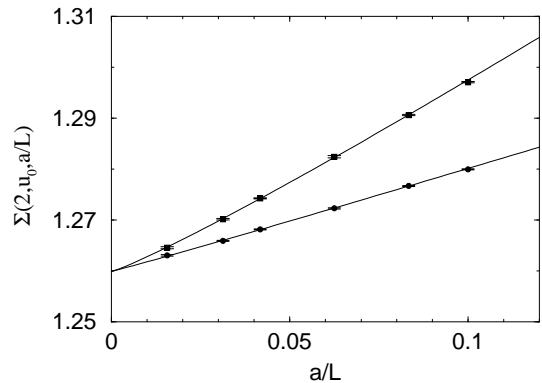}
\vspace{-0.9cm}
\caption{{}The step scaling function $\Sigma(2,u_0,a/L)$ at
$u_0=1.0595$
for two different actions.}
\label{fig:fig6}
\end{figure}

Numerical data in asymptotically free theories are standardly
extrapolated to the continuum using Symanzik's results\cite{6.3}
valid in every order of perturbation theory. Accordingly, in a bosonic
theory the leading cut-off effect is quadratic $O(a^2)$ modified by
logarithms. Results in the $O(3)$ non-linear sigma model show that
either non-perturbative effects invalidate this expectation, or the
onset of leading behavior happens at very large correlation lengths
only\cite{27.2}. In Fig.~\ref{fig:fig6} the data
obtained with two different actions
on the step scaling function
\cite{27.3} follow a linear $O(a)$ type of behavior - at least up to
correlation lengths 350. The cut-off dependence of the zero momentum
4-point function is similar\cite{27.2,27.3}.
One should be open minded.

\vspace{0.5cm}
\noindent{\bf Acknowledgements:}
I am indebted to 
Claude Bernard, Gilberto Colangelo, Tom DeGrand, Christof Gattringer,
Maarten Golterman, Anna Hasenfratz, Julius Kuti, Christian Lang, Guido
Martinelli, Steve Sharpe and my colleagues in Bern
for useful discussions. I thank for the kind hospitality during a
workshop at the Institute for Nuclear Theory, Univ. Washington, 
where part of this work was completed. 

This work has been supported in part by the Schweizerischer
Nationalfonds and the European Community's Human Potential Programme 
under contract HPRN-CT-2000-00145.

\end{document}